# Photonic Angle-of-arrival Measurement of a Microwave Signal with Range Selectivity and SNR Enhancement

Z. Cao, R. Lu, Q. Wang, A. C. F. Reniers, H.P.A. van den Boom, E. Tangdiongga, and A.M.J. Koonen

*Abstract*— A novel parallel optical delay detector (PODD) is proposed for angle-of-arrival (AOA) measurement of a microwave signal with selective measurement range and signal-to-noise (SNR) enhancement. The PODD is experimentally demonstrated by using a dual parallel Mach-Zehnder modulator. The noise robustness enhancement is experimentally investigated. Based on the proposed PODD, the measurement error for the full range can be less than 0.02 radian (0.71% of full measured range).

## I. INTRODUCTION

Determining the location of a microwave signal is of great importance for retrieving the position of objects. The parameter angle-of-arrival (AOA) or equivalently the time difference of arrival (TDOA) is required to accurately identify the position. An optical approach can benefit the measurement of AOA from several perspectives like the ultra-low loss and huge bandwidth of optical medium, which enables high accuracy and immunity to electromagnetic interferences [1-4]. Moreover, with the rapid development of ultra-low drive voltage electro-optical modulators (EOMs) [5-6] and high speed photo-diodes [7-8], barriers between electrical domain and optical domain are gradually eliminated. Recently, the parallel optical delay detector (PODD) based on one dual parallel Mach-Zehnder modulator (P-MZM) is proposed for AOA measurement [9]. In such scheme, the measurement of microwave AOA can be equivalently replaced by the phase difference (shift) measurement of optical sidebands. Its integrated parallel structure can increase robustness against environment variations due to the absence of discrete optical delay lines. However, in such system, the measurement error drastically increases when the measured phase shift (PS) moves toward $\pi$. Ideally, the optical power of the 1st sidebands will be totally suppressed when a $\pi$ PS is achieved. However, the limited extinction ratio (LER) of P-MZMs will introduce the power leakage which disturbs the final measurement results. Such disturbance (noise) is quite critical because the expected power distribution (signal) around $\pi$ is very weak. The signal to noise ratio (SNR) is thus extremely low around $\pi$. Since the LER depending on fabrication processes is difficult to eliminate, an alternative solution is to increase the measured SNR when the measured PS moves toward $\pi$. To this end, the idea is to optically transform a $\pi$ PS at the non-null point with an offset PS. In this letter, a novel PODD with range selectivity and SNR enhancement is proposed and experimentally investigated for the first time to our best knowledge. The experimental results match well with our theoretical analysis.

## II. OPERATION PRINCIPLE

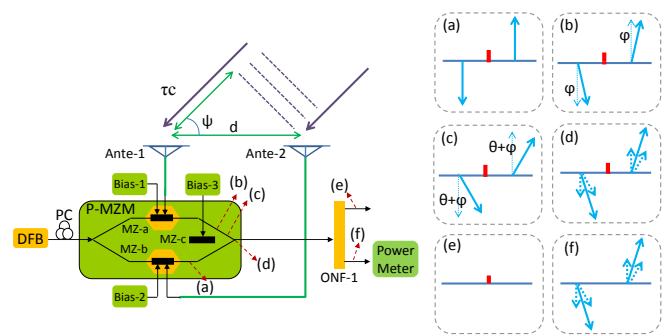

Fig. 1 The principle of proposed parallel optical delay detector (PODD) for AOA measurement.

The proposed PODD for AOA measurement is depicted schematically in Fig. 1. The distance between two antennas (Ante-1 and Ante-2) is denoted as *d*. The AOA is denoted as $\psi$ and the corresponding TDOA can be expressed by:

$$\tau = d\cos(\psi)/c \quad (1)$$

where *c* is the light velocity in air. The TDOA $\tau$ will introduce a microwave PS of $\varphi$ between Ante-1 and Ante-2 as:

$$\varphi = \tau \times \omega_m \quad (2)$$

where $\omega_m$ is the angular frequency of the microwave signal. Therefore the task of the proposed PODD is to measure the microwave PS $\varphi$ in optical domain. In Fig. 1, MZ-a and MZ-b are the sub-MZMs inside the P-MZM, MZ-c is the tunable

Manuscript received October 20, 2013. This work is supported by The Netherlands Organization for Scientific Research (NWO) under the project grant Smart Optical-Wireless In-home Communication Infrastructure (SOWICI). This work is also supported by the National Basic Research Program of China ("973" Project) (Grant No. 2012CB315701, 2012CB315702), the National Nature Science Foundation of China (Grant No. 61307070).
Z. Cao, R. Lu, Q. Wang, A. C. F. Reniers, H.P.A. van den Boom,E. Tangdiongga and A.M.J. Koonen are with the COBRA Institute, Eindhoven University of Technology, NL 5600 MB Eindhoven, The Netherlands. (E-mail: z.cao@tue.nl).
R. Lu is also with the School of Optoelectronic Information, State Key Laboratory of Electronic Thin Films and Integrated Devices, University of Electronic Science and Technology of China, Chengdu 610054, China. (E-mail: luronguo@uestc.edu.cn).

phase shifter between MZ-a and MZ-b. MZ-a and MZ-b are connected to Ante-1 and Ante-2, respectively. The lightwave from the continuous wave (CW) laser is split and modulated by two replicas of a microwave signal at MZ-a and MZ-b with spectra shown in Fig. 1(a) and (b), respectively. Both MZ-a and MZ-b are biased at the null points to suppress the optical carrier and to maximum the optical power of the 1st order sidebands. The microwave PS of $\varphi$ caused by a $\tau$ TDOA will introduce an $\varphi$ PS of the 1st order optical sidebands. The output optical signal from MZ-b with an $\varphi$ PS is then shifted with an additional $\theta$ PS at MZ-c as shown in Fig.1 (c). The output optical signals from both MZ-a and MZ-b with an $\varphi+\theta$ PS are then combined with the optical spectrum shown in Fig. 1(d). As shown in Fig. 1(e), the power of upper (or lower) the 1st order optical sidebands is measured via a power meter (PowerMeter shown in Fig.1). The optical carrier is suppressed via an optical notch filter (ONF-1 shown in Fig. 1) to remove its disturbance to the power measurement of the 1st order optical sidebands. The spectrum of the filtered optical sidebands is shown in Fig. 1(f). The electrical paths (including connections and necessary components like amplifiers) between Ante-1 and MZ-a, and between Ante-2 and MZ-b will introduce phase differences for different frequencies due to different physical lengths and impedance mismatches. Such phase differences can be easily compensated using a look-up table. According to Ref.[9], the power of the upper/lower sideband after the optical notch filter can be expressed as:

$$P_{\pm 1} = E_0^2 J_{\pm 1}^2(m)[\exp(j\varphi) + \exp(j\theta)][\exp(\mp \quad \mp \\
= E_0^2 J_{\pm 1}^2(m)[2 + \exp(j\varphi \mp \quad \mp \\
= 2E_0^2 J_{\pm 1}^2(m)[1 + \cos(\varphi \mp$$ (3)

where $E_0$ is the amplitude of the optical carrier. The modulation depth can be written as $m=\pi E_m/V_\pi$, where $E_m$ is the amplitude of arrived microwave signal, and $V_\pi$ is the switch off voltage of MZ-a/-b. $J_{\pm 1}(m)$ is the Bessel function of first kind with regard to modulation index ($m$). The high order sidebands are negligible for low driving power, which is the case for AOA measurements. To avoid measurement ambiguity, the phase item ($\varphi\pm\theta$) should be within a monotonic interval of the cosine function (e.g. [0, $\pi$]). Thus the max measurement is [$\pm\theta$, $\pi\pm\theta$]. It is obvious that the measurement range can be selected (or transformed) applying a $\theta$ offset PS. When the PS close to $\pi$ is measured, a $\theta$ offset PS with a given value can be added to prevent the null power induced measured SNR degradation. Since the power of the 1st order optical sidebands is monotonically related to the $\varphi+\theta$ PS, the value of $\varphi+\theta$ can be obtained reversely from the measured power. Since $\theta$ is adjusted with negligible loss in optical domain, there is no impact on measured results. The value required for AOA estimation is the normalized power ($P_n$), thus the value of $E_0$ and $J_{+1}(m)$ are less interesting. We can obtain the expressions of TODA ($\tau$) and AOA ($\psi$) as:

$$P_n = P_m / P_0, \quad \varphi = \arccos(P_n - 1) \pm \theta \\
\tau = \arccos(P_n - 1)/\omega_m, \quad \psi = \arccos(\tau c / d)$$ (4)

According to Eq. 4, to estimate the values of $\tau$ and $\psi$, the required parameters are $P_n$ and $\omega_m$. Since $P_m$ and $P_0$ (with zero phase shift $\varphi+\theta=0$) can both be measured and $\omega_m$ is given, the value of AOA or TDOA can be then obtained through Eq. 4 given above.

## III. EXPERIMENTAL SETUP AND RESULTS

Fig. 2 shows the proof-of-concept experimental setup for AOA (or TDOA) measurement with range selectivity and SNR enhancement. The optical carrier is generated from a DFB laser at 1550.016nm with 1dBm power. It is fed into a P-MZM after a polarization controller (PC). MZ-a and MZ-b are both biased at minimum points of their power transfer curves. Two microwave sources (LO-1 and LO-2) are synchronized to drive MZ-a and MZ-b with 12.5GHz sinuous signals. A 10MHz sinuous signal generated from LO-1 is sent to LO-2 for synchronization. MZ-c is biased at zero PS ($\theta=0$) and $\pi/2$ PS ($\theta=\pi/2$) to combine optical signals from MZ-a/-b. The optical spectrum of combined signal is shown in Fig. 2(a). The red line is for $\theta=0$ and the black for $\theta=\pi/2$, which are the same as in Fig. 2(b)-(c), and Fig. 3(a)-(c). The phase differences between LO-1 and LO-2 induced by different electrical paths and impedance mismatches are measured by a sampling oscilloscope (digital communication analyzer). It is then further calibrated via a look-up table. The output optical signal are then separated by an array waveguide grating (AWG) which acts as an optical notch filter. The channel space of the AWG is 12.5GHz and the optical signal is then separated into three channels. The optical carrier is in the middle channel (noted as CH-2) and two sidebands are in the two neighboring channels (noted as CH-1/-3). After the AWG, the power disturbance from the optical carrier is removed shown in Fig .2 (c). Its spectra is shown in Fig. 2(b) for $\theta=0$ and $\theta=\pi/2$. Optical signals from CH-1,3 are then coupled via a 3-dB coupler (OC) for convenience to show their spectra in one figure. In Fig. 2 (a) - (c), which are measured with $\varphi=0$, the power of sidebands for

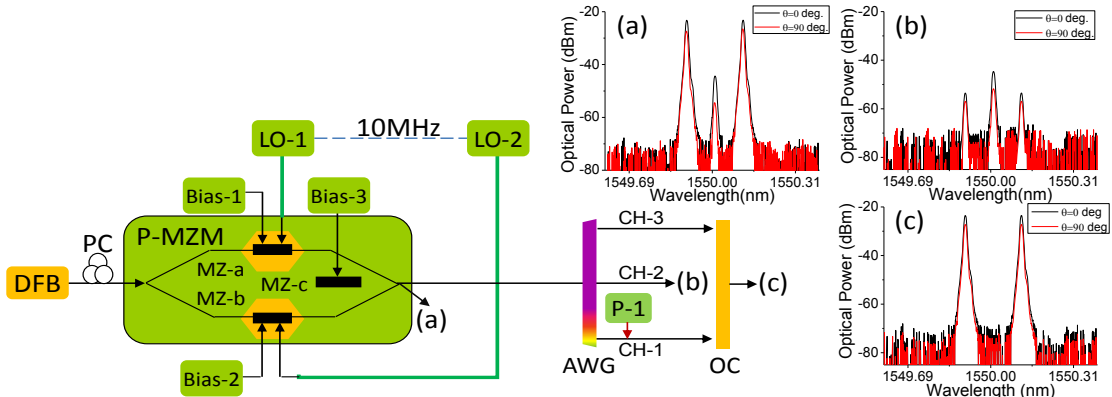

Fig. 2 The experimental setup of AOA measurement based on parallel optical delay detector, the black line for θ=0, and the red line for θ=π/2.

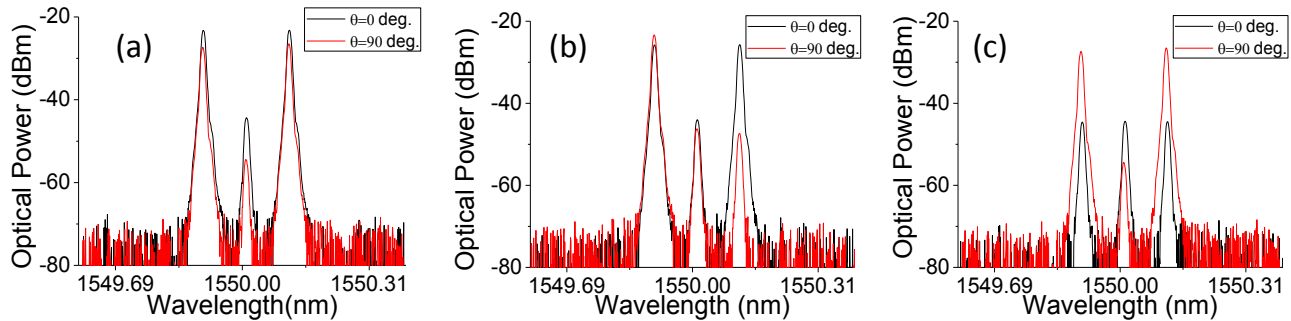

Fig. 3. The measured optical spectrum for (a) φ=0, (b) φ=π/2, (c) φ=π, the black line for θ=0, and the red line for θ=π/2.

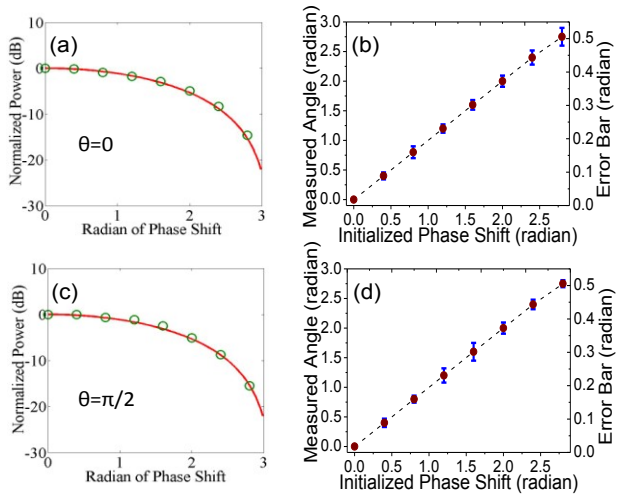

Fig. 4. (a), (c): Measured optical power (circles) and theoretical trend (curve); (b),(d): Measured phase shift (dots) and their measurement errors (vertical bars).

θ=π/2 is lower than the one for θ=0 due to the offset PS induced power decrease. The measured spectra from CH-1 with different PSs are shown in Fig. 3. Fig. 3(a) is the same as Fig. 2 (a) and is shown for comparison. For φ=π/2, with π/2 offset PS (θ=π/2), the upper sideband is suppressed due to a π PS achieved (θ+φ=π), shown as the red line in Fig. 3 (b). The lower sideband achieves its maximum with zero offset PS (θ+φ=π) shown in Fig. 3 (b). For φ=π, without offset PS (θ=0), both upper and lower sidebands are suppressed (±φ+φ=π) shown as the black line in Fig. 3 (c), while for φ=π and π/2 offset PS (θ=π/2), the power of both upper and lower sidebands are equal (±θ+φ=±π/2, cos(π/2)= cos(-π/2)) shown as the red line in Fig. 3(c). A calibration to obtain the minimum output power is carried out by optimizing the biases for MZ-a/b/c. After that, $P_0$ is obtained by measuring the maximum power of the sidebands with zero microwave phase difference between LO-1 and LO-2. In Fig. 3(a)-(c), the optical carrier is not completely suppressed mainly due to the LER, since the DC drift is mostly eliminated at the initial calibration stage.

For the AOA measurement (φ), two sets of measurements are carried out for θ=0 and θ=π/2, respectively. The measured powers versus different PS (circles) are shown in Fig. 4(a) and (c) for θ=0 and θ=π/2, respectively. The measured range is from 0 to 2.8 radian. The corresponding theoretical power distribution versus PS is also shown as the red lines in Fig. 4(a) and (c). Note that the measured value of zero PS is used for normalization and thus its measurement error is not applicable. In general, a good agreement between the theoretical analysis and measured results is obtained. However, the slight difference of the measurement error is observed. For θ=0, the measurement error increases with φ toward π. Its maximum measurement error is 0.05 radian and 1.78% of the full measured range (2.8 radian). While for θ=π/2, the maximum measurement error is 0.061 radian and 2.17% of the full measured range (2.8 radian) around φ=π/2. We can obviously see that the measurement error degrades when θ+φ is toward π as we expect. The detailed measurement errors are shown in Fig. 4(b) and (d) for θ=0 and θ=π/2, respectively. By replacing the measured results of 2.4 and 2.8 radians in Fig. 4(b) with the ones in Fig. 4(d), the measurement error can be below 0.02 (0.71% of full measured range). In a practical system, when the AOA close to π is measured, a π/2 offset PS can be added by adjusting DC bias at MZ-c.

## IV. Conclusion

We have proposed a novel photonic scheme for microwave angle-of-arrival (AOA) measurement with range selectivity and the resulting SNR enhancement based on a parallel optical delay detector. The measurement error can be reduced from 2.17% to 0.71% with the signal-to-noise enhancement. The proposed scheme can be an attractive solution for flexible and accurate angle-of-arrival measurement of a microwave signal.

**Reference**
[1] M. Jarrahi, T. H. Lee, and D. A. B. Miller, "Wideband, Low Driving Voltage Traveling-Wave Mach-Zehnder Modulator for RF Photonics," *Photonics Technology Letters, IEEE,* vol. 20, pp. 517-519, 2008.
[2] H. Huang, S. R. Nuccio, Y. Yue, J. Yang, Y. Ren, C. Wei, G. Yu, R. Dinu, D. Parekh, C. J. Chang-Hasnain, and A. E. Willner, "Broadband Modulation Performance of 100-GHz EO Polymer MZMs," *J. Lightwave Technol.,* vol. 30, pp. 3647-3652, 2012.
[3] E. Rouvalis, M. Chtioui, F. van Dijk, F. Lelarge, M. J. Fice, C. C. Renaud, G. Carpintero, and A. J. Seeds, "170 GHz uni-traveling carrier photodiodes for InP-based photonic integrated circuits," *Opt. Express,* vol. 20, pp. 20090-20095, 2012.
[4] H. Ito, S. Kodama, Y. Muramoto, T. Furuta, T. Nagatsuma, and T. Ishibashi, "High-speed and high-output InP-InGaAs unitraveling-carrier photodiodes," *Selected Topics in Quantum Electronics, IEEE Journal of,* vol. 10, pp. 709-727, 2004.
[5] R. K. Mohan, Z. W. Barber, C. Harrington, and W. R. Babbitt, "Frequency Resolved Angle and Time Difference of Arrival Estimation with Spatial Spectral Holography," 2010, p. OWF3.
[6] B. Vidal, M. Á. Piqueras, and J. Martí, "Direction-of-Arrival Estimation of Broadband Microwave Signals in Phased-Array Antennas Using Photonic Techniques," *J. Lightwave Technol.,* vol. 24, p. 2741, 2006.
[7] S. Pan, J. Fu, and J. Yao, "Photonic approach to the simultaneous measurement of the frequency, amplitude, pulse width, and time of arrival of a microwave signal," *Opt. Lett.,* vol. 37, pp. 7-9, 2012.
[8] X. Zou, W. Li, W. Pan, B. Luo, L. Yan, and J. Yao, "Photonic approach to the measurement of time-difference-of-arrival and angle-of-arrival of a microwave signal," *Opt. Lett.,* vol. 37, pp. 755-757, 2012.
[9] Z. Cao, H. P. A. van den Boom, R. Lu, Q. Wang, E. Tangdiongga, and A. M. J. Koonen, "Angle-of-Arrival Measurement of a Microwave Signal Using Parallel Optical Delay Detector," *Photonics Technology Letters, IEEE,* vol. 25, pp. 1932-1935, 2013.